\documentclass[conference]{IEEEtran}
\IEEEoverridecommandlockouts
\usepackage{cite}
\usepackage{amsmath,amssymb,amsfonts}
\usepackage{algorithmic}
\usepackage{adjustbox}
\usepackage{graphicx}
\usepackage{multirow}
\usepackage{textcomp}
\usepackage{lastpage}
\usepackage{ctable}
\usepackage{fancyhdr}

\newcommand{\rom}[1]{\uppercase\expandafter{\romannumeral #1\relax}}

\def\BibTeX{{\rm B\kern-.05em{\sc i\kern-.025em b}\kern-.08em
		T\kern-.1667em\lower.7ex\hbox{E}\kern-.125emX}}
\begin{document}
	
	\title{Destination similarity based on implicit user interest}
	\author{
		\IEEEauthorblockN{Hongliu CAO\IEEEauthorrefmark{1}, Eoin Thomas\IEEEauthorrefmark{1} }
		\IEEEauthorblockA{\IEEEauthorrefmark{1} Amadeus S.A.S., AI Research Department, 485 Route du Pin Montard. BP 69, 06902 Sophia Antipolis Cedex, France
		}
		\IEEEauthorblockA{E-mail: hongliu.cao@amadeus.com}
	
	}

	\maketitle

	\begin{abstract}
	With the digitization of travel industry, it is more and more important to understand users from their online behaviors. However, online travel industry data are more challenging to analyze due to extra sparseness, dispersed user history actions, fast change of user interest and lack of direct or indirect feedbacks. In this work, a new similarity method is proposed to measure the destination similarity in terms of implicit user interest. By comparing the proposed method to several other widely used similarity measures in recommender systems, the proposed method achieves a significant improvement on travel data.       
		
			\begin{IEEEkeywords}
		Destination similarity, Travel industry, Recommender System, Implicit user interest
	\end{IEEEkeywords}
	\end{abstract}
	\section{Introduction}
	The travel industry is more and more relying on e-commerce nowadays as online solutions have made life more convenient and comfortable for everyone. However, unlike the online shopping industry, the travel industry is more complicated to analyse in four ways: 1. user data are much more sparse. A user on amazon may have a lot of search and purchasing history on Amazon during a short period like one month. But in the travel industry, a passenger may only reserve fight tickets once a year. 2. For platforms such as Amazon, each good/item has a clear hierarchical category (e.g, diapers belong to the category of Baby care, which belongs to a higher level category of Baby), which may be used as the definition of user interest. In this application for the travel industry, we  consider the destination as the item such as Paris, London or Shanghai. But it is hard to define a category for Paris in terms of explicit user interest. 3. Account and user information is necessary for online purchasing on Amazon, which makes it easier to group the history of searches and purchases by user. For online flight bookings for example, the user does not have to create an account during the booking phase. Travelers often book flight tickets, train tickets, hotel, activities on different platforms, which makes it harder to group the purchases by user. 4. Most online products such as movies or Amazon products have ratings as user can give feedback easily. However, given that a traveler has been to Paris for example, it is hard to get a rating from the traveler on the Paris trip. Because a trip includes tickets booking, hotel booking, Point Of Interests (POIs) visits, food, etc. On the other hand,  it is difficult for travelers to rate a trip to a certain destination as there are many variables. All these differences make it a great challenge to collect, analyse and understand user data in the travel industry.      
	\iffalse 5. There are many choices (airline companies, travel agencies and meta search websites) when searching for a flight, while there are fewer choices for shopping diapers online. This means that, the search and reservation data for the same user is distributed on many different online platforms. Due to GDPR, it is impossible to share information among platforms to match the user information. \fi
	
	 One important way to help understand traveler trends is destination similarity. Destination similarity is very important for the travel industry:

	\begin{enumerate}
	    \item For travelers: we want to help travelers to find similar destinations that they may be interested in. For example, with the search or booking history of a traveler, similar destinations can be recommended to the traveler for the next trip. Another example is 
	    that the unprecedented COVID-19 crisis has introduced a number of travel restrictions which will prevent leisure travelers from reaching to their dream destination. With the destination similarity, we can recommend them alternative non-restricted destinations they might also be interested in.
	    \item For tourism offices: tourism offices can better identify their competitors for each origin market. This can allow them to better distinguish themselves and target travelers considering trips to similar locations.
	    \item For online advertising companies, destination similarity can be used to identify if the current user who is searching for destination A would be interested in their impression of destination B, to improve the click through rate or conversion rate. 
	    \item For a sustainable travel industry, the destination similarity can be used to suggest destinations that travelers might be willing to visit (so with the potential to convert a search into a booking), but closer to their home or simply better served by direct flights, thus reducing the CO2 emissions linked to transport. It can also be a solution to fight over-tourism problems by recommending similar destinations with fewer tourists or offering a more local an authentic experience, making travel even more rewarding.
	\end{enumerate}

In this work, we propose to measure destination similarity from the search logs based on anonymized cookie ID.  Various similarity measures (among users or items) have been proposed  in the literature for collaborative filtering (CF) based recommender systems. However, most of these measures are proposed based on the users' ratings, while there's no rating information for a destination (city) in the travel industry. This makes many similarity measures not suitable for our problem. To fill this gap in the literature, we investigate different possible solutions and propose a new similarity measure, which has superior performance than the state of the art methods.

The remainder of the paper is organized as follows: the background and related works are introduced in Section \rom{2}. In Section \rom{3}, the proposed similarity measures are introduced. We describe the data sets chosen in this study and provide the protocol of our experiments in Section \rom{4} and  the experimental results in Section \rom{5}. The final conclusion and future works are given in Section \rom{6}.

\section{Background and related works}
%Data nature, culture difference, 
Generally speaking, before a traveler makes a booking for a holiday, there are two preliminary steps: inspiration and search. There are a lot of information sources that can inspire travelers such as a movie scene or recommendations from relatives. During the inspiration step, the user interest is broad and general, thus we can only estimate the implicit user interest. With enough accumulated motivation, travelers will then search for more detailed information from travel websites, blogs or reviews to enrich their general picture of the destination. Then, the traveler will start to search flight and hotel information. If the prices agree with the budget, the traveler will pass to the next step: booking. Otherwise, the traveler may search for another similar destination and compare the price.

The general motivation here is that: when a user searches for travel to a destination, this action shows that the user is interested in this destination. But we don’t know what exactly the user is interested in (e.g. the museum, beach, mountains or other activities), which can be called ‘implicit user interest’. The explicit user interest is difficult and expensive to get. Many companies ask the customers directly, while others try to infer from customer shopping/spending behavior. 
However, apart from the costs, the explicit user interest may not be clear for travelers themselves either.  Because tourist attractions or POIs are not the only reason travelers get interested in a destination, it may also due to the culture (local people, food, etc.), weather, events and so on. 
Capturing implicit user interest seems  easier and more direct. When a user searches both destination A and destination B, there must be some similarities between these two destinations for this user. However, the user interest changes overtime. If the time difference between two different searches is 10 months for example, these two destinations may not be similar as the user interest may shift due to the season or travel context. Hence, limiting the time period between two different searches is important.  

The research question is: given that a user has searched for several destinations, how can we determine the destination similarity? In the literature, there are many item similarity measures that can be applied. For example, in recommender systems, the CF has become the most widely used method to recommend items for users \cite{ricci2011introduction, gazdar2020new}. The core of CF is to calculate similarities among users or items \cite{liu2014new}. Destination similarity can be seen as recommending destinations for users from the point of view of CF. 

The classic CF problem has a rating matrix. Let $R = [r_{u,i}]^{m\times n}$ be a rating matrix with $m$ users and $n$ items. Each entry $r_{ui}$ is a  rating value given by user $u$ to item $i$. There are different ranges of $r_{ui}$ in real world datasets. Among which, the range 1,2,3,4,5 is adopted in many datasets such as movie reviews and restaurant reviews. Many predefined similarity measures can be used for CF. The most commonly used similarity measure are the Pearson correlation coefficient (PCC) \cite{su2009survey} and the Cosine similarity (Cos) \cite{adomavicius2005toward}: 
\begin{equation}
\label{pcc}
PCC(u, v) = \frac{\Sigma_{i\in I}(r_{u,i} -\overline{r}_u )(r_{v,i} -\overline{r}_v)}{\sqrt{\Sigma_{i\in I}(r_{u,i} -\overline{r}_u)^2} \sqrt{\Sigma_{i\in I}(r_{v,i} -\overline{r}_v)^2}}
\end{equation}

\begin{equation}
\label{cos}
Cos(u, v) = \frac{\Sigma_{i\in I}r_{u,i}r_{v,i}}{\sqrt{\Sigma_{i\in I}r_{u,i}^2 \Sigma_{i\in I}r_{v,i} ^2}}
\end{equation}

where $I$ represents the set of common rating items by users $u$ and $v$.  $\overline{r}_u$ and $\overline{r}_v$ are the average rating value of user $u$ and $v$ respectively.

Many variants of PCC and Cos have also been proposed. In \cite{shardanand1995social}, Constrained Pearson Correlation Coefficient
(CPCC) has been proposed to take  the negative rating values into account. Other studies suggest that the number of co-rated items can impact the performance of the similarity measure. Hence, the Weighted Pearson Correlation Coefficient (WPCC) \cite{herlocker2017algorithmic} and the Sigmoid Function based Pearson Correlation Coefficient (SPCC) \cite{jamali2009trustwalker} have also been proposed.

Another widely used similarity measure is the Jaccard. Jaccard similarity coefficient is one of the popular methods for calculating the similarity between users/items  \cite{arsan2016comparison, jain2020survey}:
\begin{equation}
\label{jac}
Jaccard(u, v) = \frac{|I_u\cap I_v|}{|I_u\cup I_v|}
\end{equation}

where $I_u$ and $I_v$ are the sets of items rated by users u and v respectively.
Unlike the previous two similarity measures, the Jaccard similarity considers only the number of co-rated items between two users regardless the rating values, which seems to be suitable for our problem.

Apart from previously introduced widely used similarity measures, there are some recent advances in the literature too. In \cite{gazdar2020new}, the authors list the limitations of popular similarity measures used in CF and propose a new similarity measure by combining the percentage of non common ratings with the absolute difference of ratings. However, all these similarity measures are proposed for measuring user similarities with item ratings \cite{jain2020survey,al2018similarity}. However, this is not adapted to recommend destinations to travelers for two reasons. Firstly, there are no ratings for destinations from travelers. From search logs, we can only get binary response that if a traveler searched this destination or not. Secondly, due to the user interest change, only recent searches should be used to recommend destinations to travelers, which means that there are very few searched cities. It is difficult to measure the user similarity with little information. Hence, we need to measure the destination similarity instead of user similarity. Here, we propose a new similarity measure for items without any user ratings in the next section, and apply it to destination similarity in our experiments. 

\iffalse
The main goal of CF is to recommend a subset (topN) of the unknown items to a given user. The most popular way of realizing this for datasets with ratings is to : 1. measure user similarity; 2. predict unknown ratings. 
\fi
\section{Proposed similarity measures}
To recommend a destination to a traveler who has made one or more searches recently, we can directly measure the destination similarity and recommend destinations similar to travelers' recent searches. 

Let $R = [r_{u,i}]^{m\times n}$ be a binary matrix of with $m$ users and $n$ destinations. $r_{u,i} = 1$ means user $u$ recently searched destination $i$, while $r_{u,i} = 0$ means user $u$ didn't search destination $i$. The matrix $R$ is very sparse due to two reasons: 1. there are many destinations while each traveler only knows few of them; 2. people don't plan travel frequently.  As the matrix is binary, many commonly used CF similarity measures based on ratings are less meaningful.  

In this work, a simple and easy to understand similarity measure is proposed inspired from Random Forest Similarity (RFS) \cite{cao2019random, cao2019random1}. Random Forest (RF) classifier \cite{breiman2001random} is one of the most successful and  widely used  classifiers. A RF $\textbf{H}$ is an ensemble made up with $M$ decision trees , denoted as in Equation \eqref{e2}:
	\begin{equation}\label{e2}
	\mathbf{H}(\mathbf{X}) = \{h_k(\mathbf{X}),k=1,\dots,M\}
	\end{equation}. 
	
RFS is a similarity measure inferred from RF, which is also widely used \cite{shi2006unsupervised,cao2018improve,farhadi2015gene}.	For each tree $h_k$ in the forest $\textbf{H}$, if two different instances $\mathbf{X}_i$ and $ \mathbf{X}_j$ fall in the same terminal node,  they are considered as similar: 
	\begin{equation}\label{sk}
	RFS^{(k)}(\mathbf{X}_i, \mathbf{X}_j)=
	\begin{cases}
	1, & \text{if}\ l_k(\mathbf{X}_i) = l_k(\mathbf{X}_j)\\
	0, & \text{otherwise}
	\end{cases}
	\end{equation}
where $l_k(\mathbf{X})$ is a function that returns the leaf node of tree $k$ given input $\mathbf{X}$.
The final RFS measure $RFS^{(\mathbf{H})}$ consists in calculating the $RFS^{(k)}$ value for each tree ${h_k}$ in the forest, and to average the resulting similarity values over the $M$ trees as in Equation \eqref{simil}:
	\begin{equation}\label{simil}
	RFS^{(\mathbf{H})}(\mathbf{X}_i, \mathbf{X}_j) = \frac{1}{M}\sum_{k=1}^{M} RFS^{(k)}(\mathbf{X}_i, \mathbf{X}_j)
	\end{equation}

\textbf{Cluster Consensus Similarity ($CCS$)}: RFS is mostly designed for classification problems, which is not suitable for the destination similarity problem with only user-destination interaction matrix. However, inspired by the idea of RFS, we propose a simple method named $CCS$. For RFS, each tree is trained to give a different data partition: each leaf node groups one or several instances together. In this work, the destinations searched by each user can be seen as a cluster. The destinations in this cluster share some similarity in terms of this user's interest. With this intuition, the similarity between destination $i$ and $j$ for user $u$ can be defined as:
	\begin{equation}\label{ccs1}
	CCS^{(u)}(i, j) =
	\begin{cases}
	1, & \text{if}\ r_{u,i} = r_{u,j} = 1 \\
	0, & \text{otherwise}
	\end{cases}
	\end{equation}

Similar to RFS, the final similarity between destination $i$ and $j$ is then averaged over all users: 
	\begin{equation}\label{ccs2}
	CCS^{(\mathbf{U})}(i, j)=
	\frac{1}{m}\sum_{u=1}^{m} CCS^{(u)}(i, j)
	\end{equation}

 With the proposed $CCS$ measure, a $n\times n$ destination similarity matrix can be provided (\eqref{matrix}). 
\begin{equation}\label{matrix}
\textbf{S}^{CCS} =
	\begin{bmatrix}
		CCS^{(\mathbf{U})}(1,1) & CCS^{(\mathbf{U})}(1,2)  & \dots  & CCS^{(\mathbf{U})}(1,n) \\
		CCS^{(\mathbf{U})}(2,1) & CCS^{(\mathbf{U})}(2,2)  & \dots  & CCS^{(\mathbf{U})}(2,n) \\
		\vdots & \vdots & \vdots & \vdots  \\
		CCS^{(\mathbf{U})}(n,1) & CCS^{(\mathbf{U})}(n,2)  & \dots  & CCS^{(\mathbf{U})}(n,n)
	\end{bmatrix}
\end{equation}

In the similarity matrix, each row is a similarity vector, represents its similarity to all other destinations. To avoid recommend the searched destination, the diagonal values (the similarity value to itself) are set to 0.

\textbf{ Normed Cluster Consensus Similarity ($CCS_{norm}$)}:
The proposed $CCS$ method is simple and easy to understand. The destination similarity for each user is calculated at first to reflect this user's travel interest. Then, the similarity values of all users are averaged, which can be seen as a majority voting process.

However, there are two requirements for a similarity measure. The first one is that each similarity vector can provide a good ranking, so that it can answer which are the most similar destinations given a known destination. The second one is to provide a solution when given multiple inputs. This requires that different similarity vectors (e.g. $\textbf{S}^{CCS}_{i,}$ and $\textbf{S}^{CCS}_{j,}$) should be comparable so that simple operations such as summing make sense. 
CCS can meet the first requirement, but not the second one. Because less popular destinations have very small value, especially when the user number $m$ is very large, while popular destinations have larger value. This means that if we average two similarity vectors to find destinations similar to both given searched cities, the popular one dominates the averaged results. This means that we only focus on the destinations to the popular one and ignore the less popular one. To mitigate this effect, the $CCS_{norm}$ is proposed to re-scale each similarity vector: 
	\begin{equation}\label{ccsnorm}
	\textbf{S}^{CCS_{norm}}_{i,}=
	\frac{\textbf{S}^{CCS}_{i,}}{max(\textbf{S}^{CCS}_{i,})}
	\end{equation}
 \iffalse
where $\textbf{S}^{CCS}_{i,}$ is the vector of similarity 
 between destination i and all other destinations calculated with CCS:
	\begin{equation}\label{ccsnorm}
	\textbf{S}^{CCS}_{i,}= [CCS^{(\mathbf{U})}(i,1), CCS^{(\mathbf{U})}(i,2),...,CCS^{(\mathbf{U})}(i,n)]
	\end{equation}
\fi
\textbf{ Popularity based Cluster Consensus Similarity ($PCCS$)}:
The proposed CCS\_norm method re-scales the similarity vectors for all destinations to the same range [0,1] so that they are comparable between destinations. CCS\_norm can help to avoid the situation that popular destination dominates the final similarity values when merging with less popular destinations. However, this brings another problem. Popular destinations are searched by most people, which means there are more data and we have more confidence on the similarity vector for popular destinations. For example, if unpopular destination $i$ has been searched by 2 users among 1 million users and another unpopular destination $j$ has been searched by another 2 users, the similarity between $i$ and $j$ is 0. However, the fact can be that they are quite similar, we just don't have enough data to support their similarity. Hence, we have more confidence for the similarity vector of popular destinations. With this intuition, the $PCCS$ is proposed based on the $CCS_{norm}$: 
	\begin{equation}\label{PCCS}
	\textbf{S}^{PCCS}_{i,}=
	\frac{1}{1+e^{p_i-	\textbf{S}^{CCS_{norm}}_{i,}}}
	\end{equation}

where $p_i$ is the popularity of destination $i$, defined on $b_i$, the rank of destination $i$ based on the number of searches:
\begin{equation}\label{popularity}
	p_i= 1-w\times
	\frac{b_i}{m}
	\end{equation}

Here, $w \in (0,1)$ is a parameter to control the difference between the similarity values for popular and unpopular destinations. This allows a trade-off between putting more confidence on popular destinations ($ w > 0$) and putting same confidence for all destinations ($w = 0$). When $w$ is bigger, the popular destination vectors are more weighted. However, too big $w$ may lead to over focus on popular destinations and decrease the recommendation diversity.   

%\section{Personalized similarity}

\section{Experiments}
\subsection{Description of datasets}
In this work, we use the destination search data. In this dataset, we use  anonymized user cookie id to group the searches of the same user together. Users from 5 countries are selected due to business interest. The data contain activity related to search sessions, from which only 3 columns are preserved: the anonymized user cookie id, the searched destination location and the country from which the search was made. This last field is only used to create multiple market specific datasets to preserve cultural differences in destination preferences.    

\subsection{Protocol of experiments}
The main objective of our experiments is to find the most suited similarity measure for search logs (binary user-items interaction). Apart from comparing different measures, the cultural difference and user interest change should also be taken into consideration.   

\textbf{Culture difference}
The travel preferences of a French person, for example, may be different from those of a Chinese person. To deal with this challenge, we took country/culture into account when measuring destination similarity. The destination similarity is measured by the country from which the search was made. 

\textbf{User interest change}
For most countries, user interest can changes over time. For example, summer destinations are usually different from winter destinations. To cope with this challenge, we regularly update the destination similarity to adapt to this shift in user interest. In the previous solution, recent two months data are used to calculate the similarity matrix and the results are updated weekly. 

\textbf{Training}
For the training procedure, we have data from 5 countries. For each country, 10 time periods are selected across 2019 and 2020. The three proposed methods $CCS$, $CCS_{norm}$ and $PCCS$ are compared to 4 widely used similarity measures in the literature: Cosine similarity, Pearson similarity, Jaccard similarity and Kulsinski similarity. Like Jaccard similarity, Kulsinski similarity is also a measure for binary vectors. It is less popular, but tested in many different fields \cite{lewis2019data,smailagic2018medal,levine2017acquiring} and achieving some good results \cite{vinayan2018amritanlp}. For $PCCS$ method, different $w$ values are tested and the best $w$ is selected.

\textbf{Testing}
The similarity matrix is calculated from the 8 weeks of data, and tested on the data of the following week. During the test phase, we randomly mask one searched destination for each user, and use the rest of the searched destinations to predict the masked destination. To realize this, the average of the rest searched destinations' similarity vectors is calculated. Then, we check if the masked destination is in the list of top 5 most similar destinations.

\section{Results}
\subsection{Comparison of different similarity measures}
To compare seven similarity measures, the experiments on 5 countries data with 10 time periods for each country have been done. The means and standard deviations of the top 5 accuracy (relative) over 10 periods for each method are shown in Table \ref{table1}. 
\iffalse
		\begin{figure}[htbp]
	    \centering
	    \includegraphics[width=0.49\textwidth]{f1.PNG}
	    \caption{The experimental results of seven similarity measures on 5 countries' data. X-axis is the country. Y-axis is the average top 5 accuracy. }
	    \label{fig:f1}
	\end{figure}

\fi

\begin{table*}[htbp]
\centering
\caption{The experimental results on five countries data, with ten time periods for each country. The mean and standard deviation of top 5 accuracy (relative) is shown. The baseline is Pearson method, and the numbers in this table show the  improvement against Pearson method.}
\label{table1}
 \begin{adjustbox}{width=0.75\textwidth} 

%\begin{center} 
\begin{tabular}{|l |p{1cm}| p{1.cm}| p{1.cm}| p{1.2cm}|p{1.cm}| p{1.1cm}|p{1.1cm}|} \hline
& Pearson & Cos & Jaccard & Kulsinski & CCS & $CCS_{norm}$ & PCCS \\ 
 \hline
Country1 &
 $baseline$ &
 $2.22\% \pm 0.80$ &
 $2.90\% \pm 1.13$ &
 $9.38\% \pm 2.06$ &
 $9.39\% \pm 1.84$ &
 $9.38\% \pm 1.97$ &
 $\mathbf{9.89}\% \pm 1.82$
 \vspace*{0.0mm} \\
Country2 &
 $baseline$ &
 $2.10\% \pm 1.21$ &
 $2.49\% \pm 1.57$ &
 $4.39\% \pm 1.86$ &
 $4.24\% \pm 1.93$ &
 $5.09\% \pm 1.81$ &
 $\mathbf{5.78}\% \pm 1.90$
 \vspace*{0.0mm} \\
Country3 &
 $baseline$ &
 $1.36\% \pm 0.28$ &
 $2.29\% \pm 0.75$ &
 $4.70\% \pm 1.39$ &
 $4.51\% \pm 1.36$ &
 $4.96\% \pm 1.29$ &
 $\mathbf{5.71}\% \pm 1.36$
 \vspace*{0.0mm} \\
Country4 &
 $baseline$ &
 $2.92\% \pm 0.42$ &
 $2.81\% \pm 0.53$ &
 $5.14\% \pm 0.51$ &
 $5.02\% \pm 0.48$ &
 $5.80\% \pm 0.51$ &
 $\mathbf{6.47}\% \pm 0.48$
 \vspace*{0.0mm} \\
Country5 &
 $baseline$ &
 $4.22\% \pm 0.80$ &
 $3.90\% \pm 0.77$ &
 $5.81\% \pm 0.48$ &
 $5.55\% \pm 0.45$ &
 $6.32\% \pm 0.41$ &
 $\mathbf{7.18}\% \pm 0.44$
 \vspace*{0.0mm} \\
 \hline 

Avg rank
&7.00
&5.60
&5.40
&3.20
&3.60
&2.20
&1.00
\\ \hline
Avg improvement
&0.00
&2.56\%
&2.88\%
&5.89\%
&5.74\%
&6.31\%
&7.01\%
 \vspace*{0.5mm} \\ \hline
 \end{tabular}
 %\end{center}
 \end{adjustbox}
 \end{table*}
 
The results illustrate that among all five countries, $PCCS$ always has the best performance, followed by $CCS_{norm}$. $Pearson$ is always the worst performing method among all similarity measures. Table \ref{table1} gives more details on comparison. On average, the proposed $PCCS$ improves the performance of $Pearson$ by 7.01\%. Compared to the previous selected method $Cos$, $PCCS$ gains an improvement of 4.44\% on average.

 From the results of average ranking in Table \ref{table1}, it can be observed that two best performing methods are $PCCS$ and $CCS_{norm}$, while the widely used similarity measures $Cos$ and $Pearson$ are the worst. One reason is that these two measures are not designed for the comparison for binary vectors. 
 
\begin{figure}[htbp]
	    \centering
	    \includegraphics[width=0.49\textwidth]{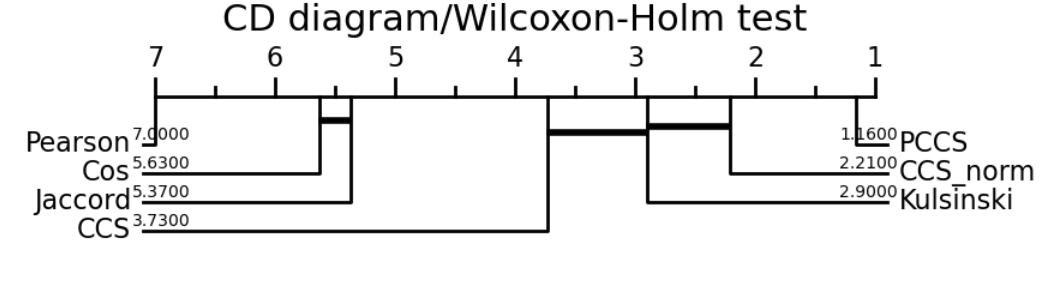}
	    \caption{The CD diagram according to the Wilcoxon-Holm post hoc test result when alpha is 0.00001. Methods connected with bold black line have no significant difference. }
	    \label{fig:cd}
	\end{figure}

\iffalse
From Table \ref{table1}, it can be seen that Country2, Country3, Country4 and Country5 have similar average performances for each similarity measure (e.g., $PCCS$ has around 54.30\% accuracy on all these four countries). However, the performance on RU is obviously worse (e.g.,  $PCCS$ has 48.67\% accuracy on RU). One possible reason is that compared to GB, FR, ES and DE, we have fewer user data from RU. For RU, $PCCS$ even  has an improvement of 9.89\% over $Pearson$ and an improvement of 7.67\% over $Cos$. Compared to the performance gap on other countries,  $PCCS$ has bigger improvement on RU, which may indicate that $PCCS$ is more suited to small data problems than other similarity measures.  
\fi
The Wilcoxon-Holm post hoc test with Critical Differences (CD) is also done to have an overall statistical comparison. The statistical test result is shown in Figure \ref{fig:cd}. Generally speaking, all three $CCS$ based methods are significantly better than $Pearson$ and $Cos$.  The proposed $PCCS$ is significantly better than all other 6 similarity measures. 
\subsection{Comparison of different training size}
\begin{table*}[htbp]
\centering
\caption{The experimental results with four weeks of training data instead of eight weeks. The mean and standard deviation of top 5 accuracy (relative) is shown. The baseline is Pearson method, and the numbers in this table show the  improvement against Pearson method.}
\label{table2}
 \begin{adjustbox}{width=0.75\textwidth} 

%\begin{center} 
\begin{tabular}{|l |p{1cm}| p{1.cm}| p{1.cm}| p{1.2cm}|p{1.cm}| p{1.1cm}|p{1.1cm}|} \hline
  & Pearson & Cos & Jaccard & Kulsinski & CCS & $CCS_{norm}$ & PCCS \\ 
 \hline
Country1 &
 $baseline$ &
 $2.21\% \pm 1.22$ &
 $2.89\% \pm 1.50$ &
 $8.88\% \pm 1.51$ &
 $8.81\% \pm 1.43$ &
 $8.87\% \pm 2.46$ &
 $\mathbf{9.45}\% \pm 2.04$
 \vspace*{0.0mm} \\
Country2 &
 $baseline$ &
 $1.91\% \pm 0.70$ &
 $2.54\% \pm 0.90$ &
 $4.37\% \pm 1.41$ &
 $4.29\% \pm 1.38$ &
 $4.99\% \pm 1.25$ &
 $\mathbf{5.70}\% \pm 1.34$
 \vspace*{0.0mm} \\
Country3 &
 $baseline$ &
 $1.20\% \pm 0.29$ &
 $2.26\% \pm 0.80$ &
 $5.01\% \pm 1.41$ &
 $4.99\% \pm 1.40$ &
 $5.31\% \pm 1.36$ &
 $\mathbf{6.01}\% \pm 1.40$
 \vspace*{0.0mm} \\
Country4 &
 $baseline$ &
 $2.85\% \pm 0.36$ &
 $2.86\% \pm 0.59$ &
 $5.62\% \pm 0.73$ &
 $5.52\% \pm 0.74$ &
 $6.24\% \pm 0.74$ &
 $\mathbf{6.90}\% \pm 0.70$
 \vspace*{0.0mm} \\
Country5 &
 $baseline$ &
 $3.79\% \pm 0.94$ &
 $3.87\% \pm 0.93$ &
 $5.97\% \pm 0.64$ &
 $5.76\% \pm 0.66$ &
 $6.40\% \pm 0.58$ &
 $\mathbf{7.26}\% \pm 0.66$
 \vspace*{0.0mm} \\
 \hline 
Avg rank
&7.00
&6.00
&5.00
&2.80
&4.00
&2.20
&1.00
\\ \hline
Avg improvement
&0.00
&2.39\%
&2.88\%
&5.97\%
&5.87\%
&6.36\%
&7.07\%
 \vspace*{0.5mm} \\ \hline
 \end{tabular}
 %\end{center}
 \end{adjustbox}
 \end{table*}
 
 In the previous section, all the experiments use the previous eight weeks of data as training data for the following week of test data.
 However, using eight weeks of training data for each market and updating the results weekly can be very time consuming and computationally expensive.  
 In this section, we try to reduce the training data to four weeks only to see to which extend the prediction performance is affected.
 
 The experimental results are presented in Table \ref{table2}. Similar to the analysis in the previous section using eight week training data, the results trained on only four weeks data also show that the proposed $PCCS$ is the best performing method while $Cos$ and $Pearson$ are the worst. On average, $PCCS$ increases the performance of $Pearson$ by 7.06\% and increases the performance of $Cos$ by 4.67\%, which is also similar to the conclusion in the previous section.  
 
 However, compared to the results in Table \ref{table1}, the average performance of each similarity measure is not strongly impacted by the reduction of training data size. The method that has the biggest difference is $Cos$, with a reduction of 0.65\% of accuracy. $PCCS$ has a performance reduction of 0.42\% on average. But when we look into each country, it can be found that the differences on Country2, Country3, Country4 and Country5 are negligible, but the performance on Country1 has a drop around 1.76\% (this analysis is based on the comparison of the absolute performances, which is not disclosed).  One possible reason can be that, the data coverage in Country1 is not very good and the data size is much smaller than other countries, which can also explain this big performance drop on Country1 than other countries. 
 
 The objective of this experiment is to answer the research question: how many data are enough to have a good prediction performance but at the same time keep the computational efficiency. The experimental results show that for countries with good data coverage, four weeks training data are enough compared to eight weeks training data. However, for countries without a proper data coverage, it's better to use eight weeks training data.  
 
 \iffalse
 \subsection{Comparison of updating frequency}
In the previous section, one possibility of improving the computational efficiency has been discussed in terms of reducing the training data size. Another way to reduce the computational cost is to update the results less frequently.  
 \fi
\subsection{Comparison of data from 2019 and 2020}
Due to the crisis of covid-19, the data volume of  2020 is much smaller than the volume of 2019 data.  The question in this section is: should we use data from 2019 or 2020 as training data for the prediction of  2020?

To answer this question, we choose the first week of June 2020 as the test data and use the data from April and May 2019 and 2020 respectively as training data to compare their prediction performance. We choose this time period is because there were confinement in Europe during April and May 2020, and the data volume is lower compared to the same period in 2019.  

\begin{figure}[htbp]
	    \centering
	    \includegraphics[width=0.45\textwidth]{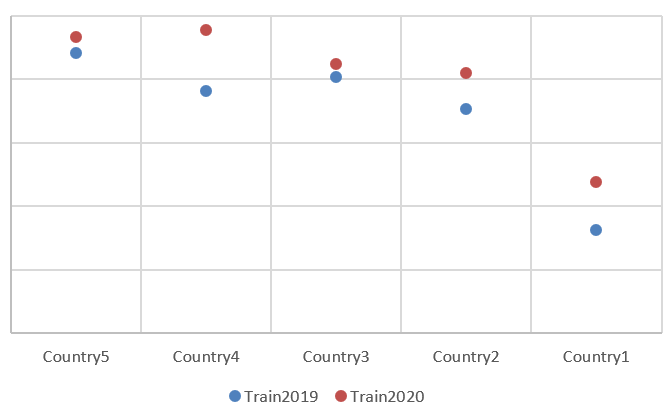}
	    \caption{The comparison of the prediction performance between 2019 data and 2020 data. X-axis is the country, Y-axis is the accuracy.}
	    \label{fig:vs}
	\end{figure}

The experimental results are shown in Figure \ref{fig:vs}. Globally speaking, 2020 data have better perdition performance than 2019 data even though the 2020 data volume is much smaller. The smallest difference happens in Country3 with 1.05\% accuracy gap, while the biggest difference happens in Country4 with 4.75\% performance gap.  One possible reason may be the change of user interest. Another possible reason may be that the covid-19 crisis has changed user's behavior (e.g. more local travel than international travel). 

\section{Conclusion}
In this work, we have presented the challenges of recommending destinations in the travel industry and the differences of recommending items other e-commerce sectors. The challenges of extra sparseness, dispersed user history,  change of user interest, few direct or indirect feedbacks  make it much harder to understand travelers and make recommendations more difficult than for other online consumers. 

To tackle these challenges and to understand travelers, we decide to measure the destination similarity in terms of traveler's implicit user interest. There are many similarity measures proposed in the field of collaborative filtering. However, most of these measures are designed for user-item interaction with ratings. Hence we propose a new similarity measure for user-item interaction without ratings to deal with the challenges in travel industry. The proposed $PCCS$ is inspired from Random Forest Similarity to take user interest into account. The destination popularity is added to adjust the magnitude of each similarity vector so that the similarity vectors of different destinations can be fused correctly. After comparing seven different similarity measures on real world data, the proposed $PCCS$ is proved to be the best solution.

However, there are some improvements can be done. Firstly, we can expand single-source to multi-sources.  Users can be limited by the knowledge of destinations, which means that there are destinations users never search because they don't know them not because they are not interested in them. Search logs can only reflect the similarity among destinations known to users.  In this way, apart from the implicit user interest from search logs, other sources can be added such as destination images or descriptions.  By using multi-source data, the implicit user interest can be fused with the destination information to provide a more meaningful similarity measure. Secondly, more user information and session information can be collected to provide a more personalized similarity measure. But more input information may also limit its use in real world use cases. 
\iffalse
Thirdly, a unsymmetrical similarity can be proposed instead of a symmetrical one. Most similarity or distance measures are symmetrical due to the thinking that: if A is similar to B, B is similar to A. \fi

	\bibliographystyle{ieeetr}
	\bibliography{sample}
	
\end{document}